# Effective System for Pregnant Women using Mobile GIS

Ayad Ghany Ismaeel
Department of Information System Engineering- Erbil Technical College- Hawler Polytechnic University (previous FTE- Erbil), Iraq.

Emad Khadhm Jabar
Department of Computer Science-University of Technology, Baghdad-Iraq.

## ABSTRACT
World Health Organization showed at one year about 287 000 women died most of them during and following pregnancy and childbirth in Africa and south Asia. This paper suggests mHealth system for serving pregnant women, that proposed system is first an effective mHealth system works base on mobile GIS to select adjacent care centre or hospital maternity on Google map at online registration for woman pregnant, that is done when the pregnant woman will send SMS via GPRS network contains her ID and coordinates (Longitude and Latitude) the server when receive it will search database support that system and using the same infrastructure for help the pregnant women at her location (home, market, etc) in emergency cases when the woman send SMS contains her coordinates for succoring. Implement the proposed pregnant women system shows more effective from view of cost than other systems because it works in economic (SMS) mode and from view of serve the system can easy and rapidly manage when achieving locally registration, succoring in emergency cases, change the review date of pregnant woman, as well as different types of advising.

## 1. INTRODUCTION
World Health Organization says about 800 women die from complications of pregnancy or childbirth-related worldwide at day. 99% of maternal deaths in poor countries and only 46% of their women benefit from skilled care during childbirth, i.e. which not help the millions of births a midwife, doctor etc. The women in rich countries obtain at least four visits prenatal care, rounded by a skilled health worker during childbirth and access to post-natal care while in poor countries just over a third of all pregnant women have recommended four visits prenatal care. The important factors that prevent women to receive care during pregnancy are poverty and distance [1].

Solving this problem sure will think about mHealth system for pregnant women to succor them, not like these traditional mHealth systems which are coming as corollary of increasing using mobile telephone, this advantage subscribers are taken in the country to improve the health of their citizens and overcome existing communication barriers by broadcasting SMS text messages to all mobile telephone for the pregnant women via pregnancy care advice as shown in Figure 1 [2].

## 2. RELATED WORK
Bangladesh experience [2010] allows the mobile users at 2010 subscribing, at a reduced rate, to SMS service that broadcasts messages in health topics. Health workers in communities throughout the country can advise Bangladesh as case study via SMS the patients through their mobile telephones. From those patients the pregnant woman can register their mobile numbers to receive prenatal advice [2].

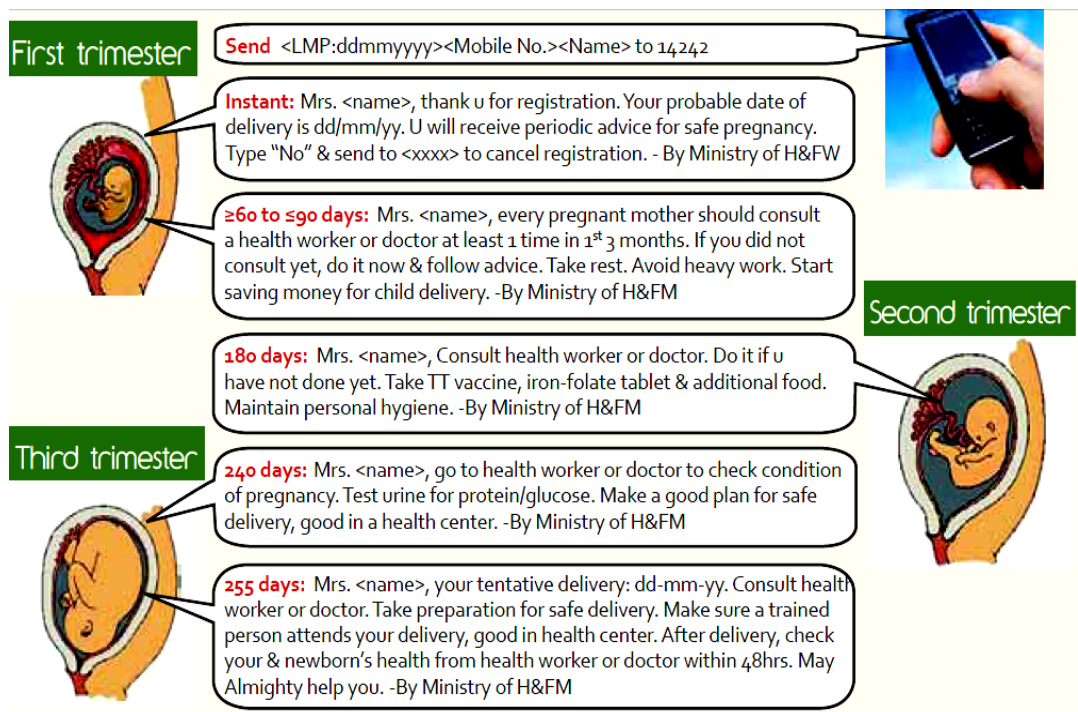

**Fig. 1:** Shows an example of traditional mHealth system, which focuses on registration and advising services [2].

Millennia2015 [2010] Under supervision of UNESCO the Women and eHealth WeHealth as international working group was released at Aug 2010 in Namur, WeHealth supports actively women with technologies to access a better healthcare, the benefits of that system are [3]:

1) Increase use and access to health services
2) Low the maternal morbidity and deaths
3) Minimize death of infants at birth.

V. M. Pomazan, Lucian Petcu and others [2009] designing eHealth monitoring system to the data transmittance and processing, they referring to be implemented in an eHealth monitoring system dedicated to pregnant women with chronically diseases. The suggested design will transfer data from different standards to make the medical staff enable to update the patient's data as well as increase the efficiency of disease management [4].

The whole previous eHealth/mHealth systems for pregnant women are using or offers advising to use mobile phone in their services but each of them has weaknesses like there isn't registration using mobile call for review, changing the date of next review, using mobile to succoring in emergency cases, etc and to overcome on the problems above, there is needed to mHealth system for pregnant women can overcome on these weaknesses, the suggested mHealth system must be based on the following techniques and modes [5]:

*A.* Mobile GIS: integrates one or more of the following technologies (mobile devices, Global Position System GPS and wireless communications for Internet GIS access).
*B.* System of mHealth works using GPRS network, and
*C.* Short Message Service SMS.

## 3. MOTIVATION

The aim of this paper reached to mHealth system for pregnant women can cover additional to advice, achieving easy registration (from their homes) in maternity care center or specialty hospital closest, changing the date of next review for pregnant woman and when this woman needed a succoring in emergency cases wherever her location the scouring facility (car, helicopter, boat life) can reach to her and in the type which is needed because the succor for blood pressure different from diabetes, heart, asthma, etc, so the proposed mHealth system will involve the following characteristics [5]:

A. The Mobile GIS Technique:
The proposed system will base on supporting of a mobile build-in GPS technique on devices like iPhone, Windows phone, iPad, etc.
B. Modes of transmission as follow:
 1) GPRS mode: The mobile sends data through GPRS data channel to special TCP/IP server linked to the Internet, or a PC with fixed Internet IP address, and
 2) SMS mode: The mobile sends data through SMS to the receiving terminal, comparing to the modem solution; the SMS solution with mobile GIS is more economical.
C. Server of TCP/IP mode:
There are a fixed and dynamic IP address, the user require here reconfigure the IP address setting of the mobile terminal to align it to any of the users desired output this remote setting makes the connection easy.

## 4. ARCHITECTURE OF PROPOSED PREGNANT WOMEN SYSTEM

The main tasks of suggested design (mHealth) for pregnant women system is summarized at flowchart shown in Figure 2:

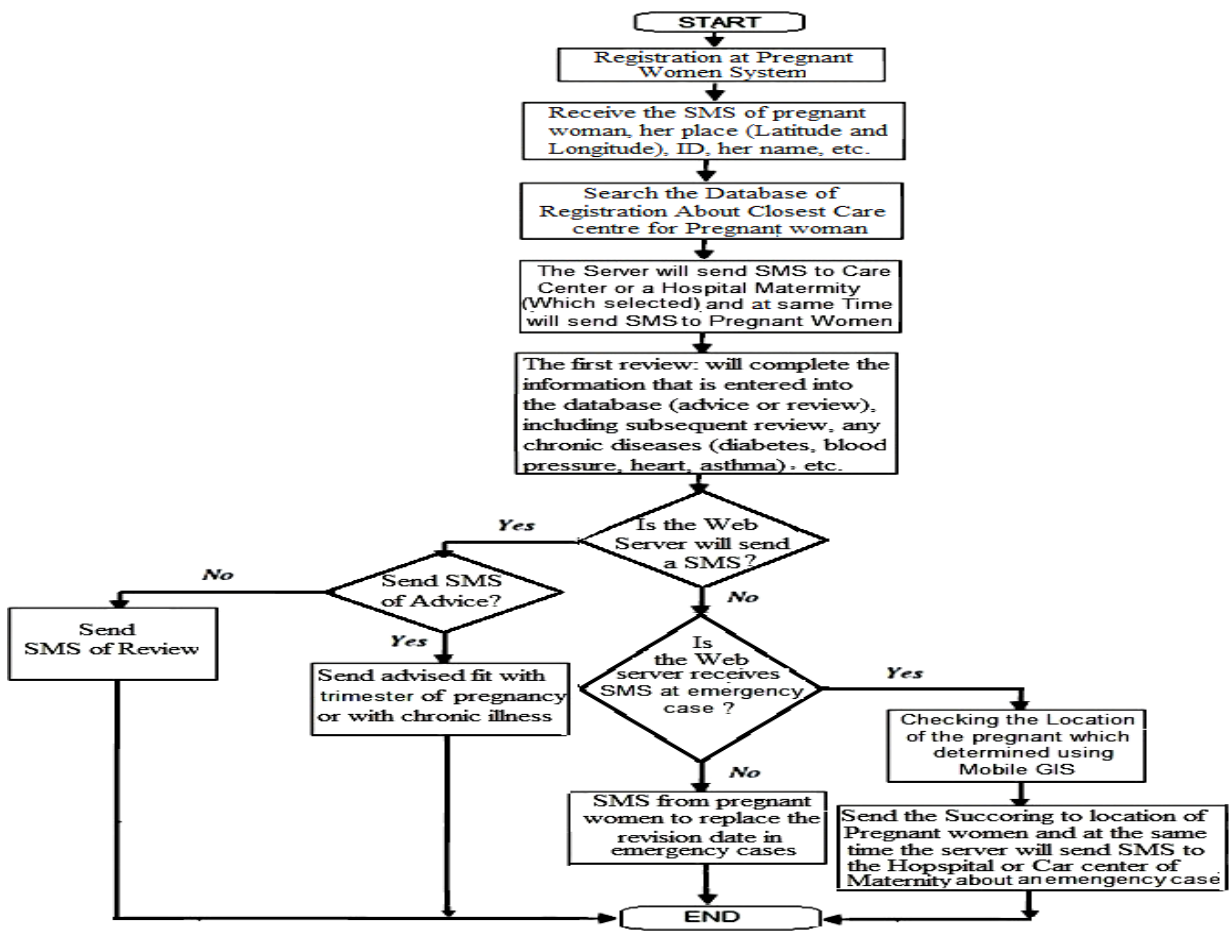

**Fig. 2: Reveals flowchart for the main tasks of proposed pregnant women system.**

Figure 3 shows a diagram of general tasks of pregnant women system which is done in two ways from woman to server (way of sending) and vice versa from server to woman (way of receiving). In the first way the woman send SMS via GPRS network for registration, changing the date of review, or request of succoring in emergency case, while at the second way may the woman receive SMS cross GPRS network for advising or calling for review. The DM which is make checkup for woman in the review will enter the review information using web interface created in proposed system for that reason.

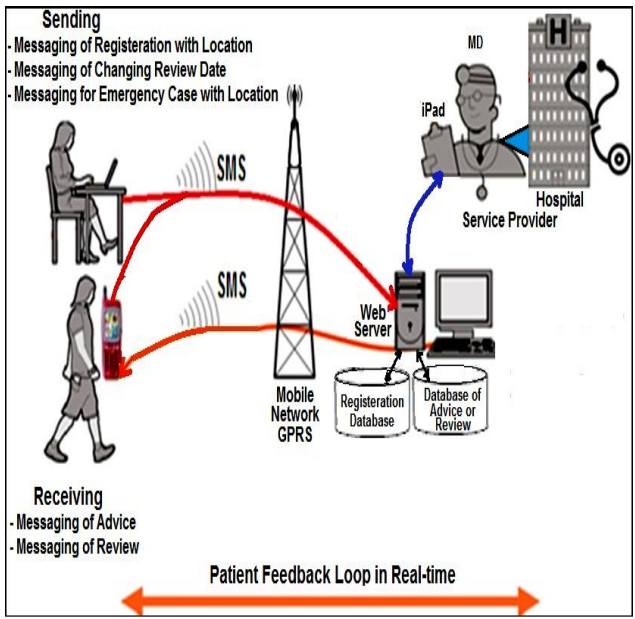

**Fig. 3: Shows diagram of proposed mHealth system for pregnant women.**

The architecture of suggested system for pregnant women involves four modules to cover the tasks of that system as shown in Figure 4, these modules are:

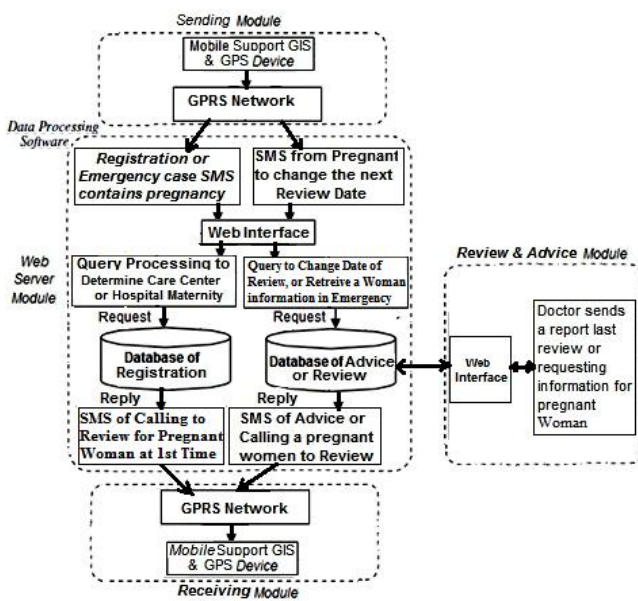

**Fig. 4: Shows architecture of suggested system for pregnant women involves four modules.**

A. Sending module:
In this module any pregnant woman needed to be serve by proposed mHealth system can use her mobile which supported GPS technique (built-in) to locate her coordinates (Latitude and Longitude) then sending SMS for multiple reasons which are:

1) Registration: at first time must be register via web interface (No duplicate), without this registration the system can't recognize the pregnant woman which needs advice, revision, succoring, etc. when the SMS reach from pregnant woman contains her location (Latitude and Longitude), ID (phone number), name, and age to the server. That server will search the *registration* database which contains table called *info* that table involve the location (Latitude and Longitude) of all care centers or hospitals maternity in the zone and the capacity of each center or hospital (The largest number of women can receive them in the care center or hospital) as shown in Table 1.

**Table 1: Fields of info table at registration database.**

| Field | Type |
|---|---|
| Maternitycarecenter-Id | Int(10) |
| Maternitycarecenter-name | Varchar (20) |
| Zone | Varchar (7) |
| Latitude | Varchar (10) |
| Longitude | Varchar (10) |
| Noofregister-women | Int(5) |
| Capacity | Int(5) |

And based on the info table the server will select the closest care center or hospital maternity for pregnant woman using the following algorithm.

---
**Algorithm of Selecting a Maternity Care Center**
- Start at source node (Location of pregnant woman)
- Move outward
- At each adjacent/neighbor u (care center of pregnant women)
  - Find node *u* that
    - *Is u adjacent to source; and*
    - *How many women registers at u*
  - Compute:
    - *The distances(the expecting directions) from source to each neighbor u*
    - *Select the shortest distance, between source and u*
---

Then the server will send SMS for pregnant woman about the selected care center or hospital maternity and at the same time create a record for this pregnant woman at advice or review database.

2) An emergency case: the woman can send SMS with her location (Latitude and Longitude) from her mobile which is support GPS technique to the server of system, i.e. call for succoring/help in an emergency case and the server will send succoring facility (car, boat life, or helicopter) and with type which is needed the pregnant woman to her location.

3) Change the date of review: the woman can change the date of review by send SMS to server when she can't come for any reason.

B. Receiving module:
The pregnant woman in this module wills receive SMS from server of proposed system as follow:

1) Advice: the server will send advice based on the trimester of pregnant woman, i.e. base on the trimester arrangement (1st, 2nd or 3rd).
2) Call for review: the server will send SMS to pregnant woman at enough time before date of review, and the server done that based on field of (next review) on review record which is created by server at each review.

C. Review and advice module:
This module works in two parts at any review for pregnant woman to the care center or hospital maternity the server of system will transport automatically the initial information like the name of pregnant woman, her phone number, location (Latitude and Longitude), age, etc which are received by SMS of registration while the other data of record will complete by the MD's iPad (Doctor which making checkup for pregnant woman), so the data for each new record of review table at advice or review database shown in Figure 5.

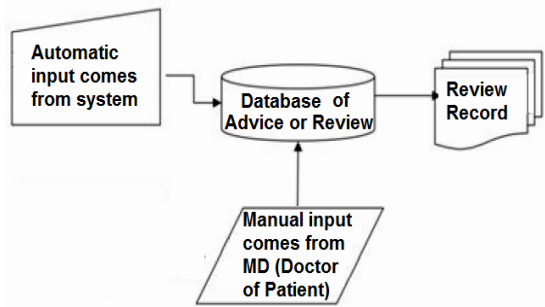

**Fig. 5: Shows the ways of input to review table at advice or review database.**

Another important table at *advice or review* database is *advice* table which contains multiple fields as shown in Table 2.

**Table 2: Fields of *advice* table at *advice or review* database**

| Field | Type | Content as Example |
|---|---|---|
| ID-code | Int(10) | 25502131 |
| Phone-of-woman | Varchar(20) | 07504432147 |
| Trimester-No | Varchar(7) | 1st, 2nd, or 3rd |
| Advice | Yes/No | (Yes) if the advice done else (No) |
| Type-of-advice | Varchar(10) | (Normal) if server done that else (other) |
| Who-advisement | Varchar(10) | Server, MD, or Admin |
| Message | Varchar(250) | SMS contains advice as shown in Fig 1, other advice SMS done by MD, or Administrator at necessary case |

D. Web Server module:
That module related with server tasks, which represented by sending and receiving SMS from and to pregnant woman, as well as transport the initial data to review record at *advice or review* database and search the registration database about the closest care center or hospital maternity for pregnant woman etc. admin server of mHealth pregnant women system can advising those women about food, chronic disease, sport, etc on the website based on set of senior doctors (committee of advising) or sending SMS when the case require that.

## 5. EXPERIMENTAL RESULTS
Implementing the design of mHealth system reveals below:

### 5.1 The Requirements
The configuration of suggested system can be divided into::

1) Software: needed Windows server 2008 to setup and install web server, the other important software C# under Visual studio 2010 package which is used in the web interfaces for proposed system and to connect between the web server and databases, then for all tasks which are seen in modules of the proposed system, last software needed MySQL package use to construct the databases and their tables and the important one the *review* table which is created by server at any new review for pregnant woman and the GUI which based on that table and its fields shown in Figure 6, that web interface is filling by MD who is checkup the pregnant woman.

**Fig. 6: Shows GUI (record) of *review* table at *advice or review* database for certain woman.**

2) Hardware: will need server can support Windows and Microsoft packages like HP server, a mobile can support GPS technique (built-in), here selected windows phone which is friendly with Microsoft packages (the compatibility makes avoid conflict) and cheap comparing to iPhone, iPad, additional will need a resource of Internet Service Provider ISP to connect via GPRS network.

### 5.2 Implement of proposed pregnant system
The proposed system is implemented at Erbil city and focus on the important tasks of system, like the registration using web interface from any PC connected to internet or from her mobile for pregnant woman directly.

When the server receiving the registration's SMS from pregnant woman as shown in Figure 7, that server will search the *info* table at *registration* database to find at 1st time all care centers or hospitals maternity closest to the location of pregnant woman.

Then the server selecting the 3 closest positions in distances (which are closet to source, i.e. pregnant woman) using the algorithm of selecting a maternity care centre as referring to in subsection 4-A; 1, at the same time must be checking these 3 position which one can accept the pregnant woman (under its capacity). Figure 8 shows example of pregnant woman location as source on Google map and 3 care centers and hospital which are closet to position of pregnant woman which are selected by server.

Table 3 shows the distance between each position and the source (in yellow color), and how the server selected maternity hospital (C) as care centre although the Ankawa (A) which is closet one to source because (A) isn't under capacity, so the server will leave Ankawa and select maternity hospital.

**Table 3: Shows computation the algorithm of selecting**

| Case | Position | Distance km | Under Capacity |
|---|---|---|---|
| 1 | Source to A (Ankawa) | 0.5 | No |
| 2 | Source to C ( Maternity Hospital ) | 3.7 | yes |
| 3 | Source to B (Tayrawa) | 6.5 | yes |

The pregnant woman received SMS from server at the same time will create new record on *review* table onto advice *or review* database, the server at the 1st review added to that record prime information of pregnant woman which obtained from registration SMS in addition the care centre or hospital maternity which selected for her while at the next reviews will add the prime information of pregnant woman from previous review and other information of review will add to record manually by MD. The proposed system checking *review* table onto advice *or review* database before the date of next review (before 3-7 days) and that done by search the last review record of pregnant woman specifically. The server will check the field of next review on the review record and send SMS to the pregnant woman for review, when the pregnant woman receives the SMS of review must be confirm or change the date of review (cases of acceptable excuse only) to suitable date as shown in Figure 9.

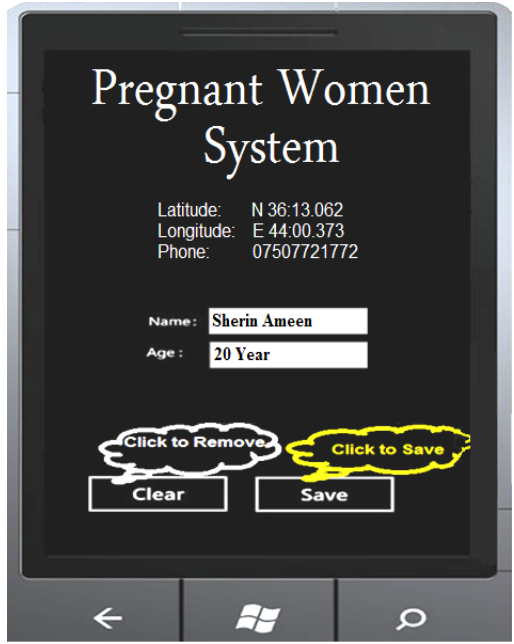

**Fig. 7: Reveals registration of pregnant woman at proposed system using her mobile.**

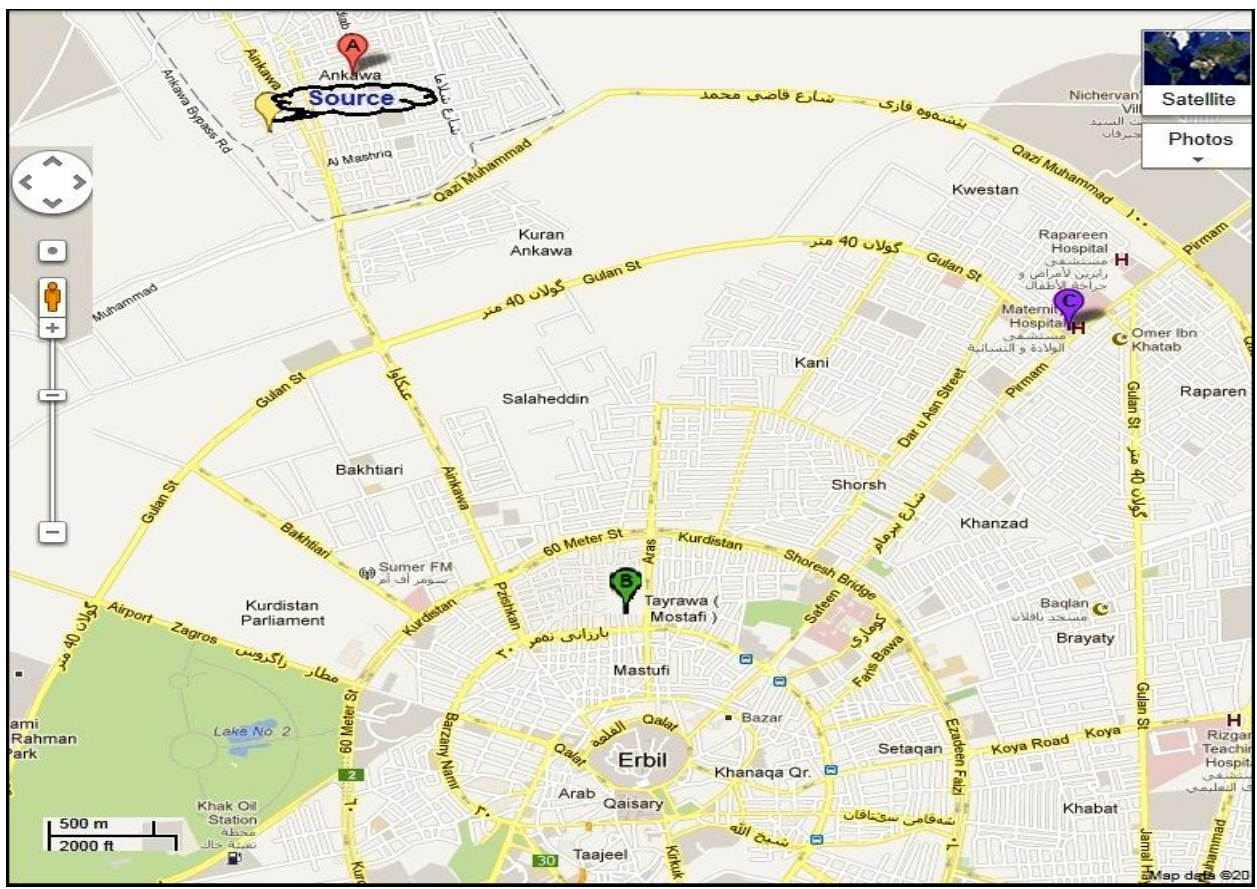

**Fig. 8: Shows the result of seeking (info table), and the server will select 3 closed places (care centre) to location of pregnant woman (source).**

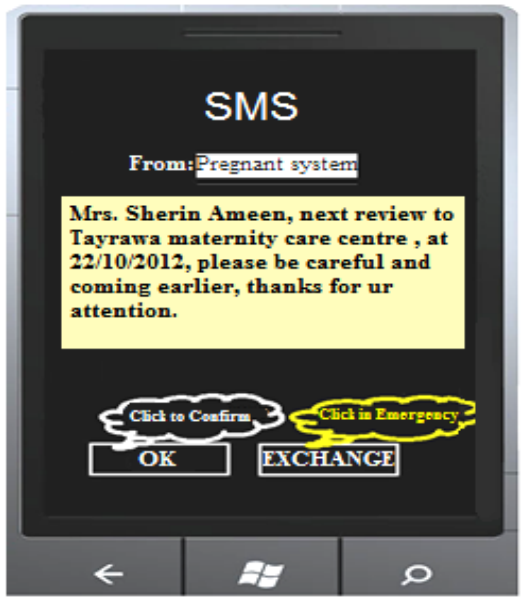

**Fig. 9: Reveals a call for review SMS received by pregnant woman.**

Another SMS may receive the pregnant woman related about advice this done by server directly (when check the field content of the trimester of pregnant is it 1st, 2nd, or 3rd), MD or consult (group of seiner doctors which are supervision of proposed mHealth system) that done by administrator of of proposed system. Figure 10 shows the advice which is done by server for pregnant woman at the 3rd trimester.

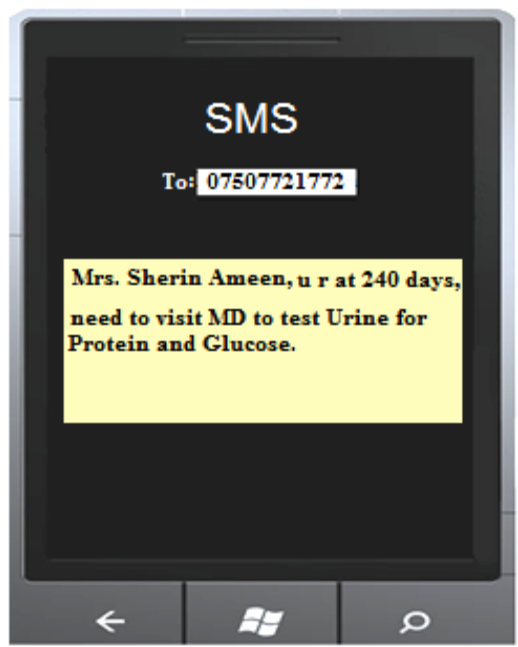

**Fig. 10: Shows advice SMS for pregnant woman.**

In, the proposed mHealth system the pregnant woman which was registered before she could sent SMS at emergency cases to the web server contains her ID and coordinates (Longitude and Latitude) via GPRS network to obtain succoring/help. Then the server will locate the woman on Google map and retrieve the pregnant woman's information from the review table in advice or review database, and based on these information will send succoring facility (car, life boat, or helicopter) to this pregnant woman.

## 5.3 Discussion of the results
Compare the proposed pregnant women system with other related systems shown in Table 4.

**Table 4: Comparing the proposed system with other systems**

| Feature | Proposed System | Bangladesh experience [2] | Millennia2015 [3] |
|---|---|---|---|
| Using Mobile GIS | Yes | No | No |
| With mobile GIS can obtain additional services (at home of pregnant woman or at her place) | Selecting closest care centre or hospital maternity at registration, obtain succor in emergency at location of pregnant woman, reached to location of pregnant woman for vaccination, etc | There isn't | There isn't |
| Using SMS from pregnant women to server and vice versa | Yes | Yes, but from pregnant women to server only | No |
| Interactive via web interface | Yes | No | Somewhat |
| Rapidly and easily obtain services | Yes, because it use mobile GIS | Can't reach to level of proposed system | No |

## 6. CONCLUSIONS AND FUTURE WORK

### 6.1 Conclusions
The important conclusions which are obtained from proposed pregnant women system shown below:

1) The proposed system is more effective than other systems because it support locally registration of pregnant woman (from her home or her place), call for review, advising, interactive, etc as referring to in Table 4.

2) Proposed system based on mobile GIS satisfying online , rapidly and easily services like registration of pregnant woman, call for next review, advice, etc using SMS mode, which is economic relatively (effective cost) comparing to other modes like modem solution.

3) This proposed system can offer succoring locally in an emergency cases for pregnant women which is registered, e.g. when the pregnant woman at home, market, or any other location and require succor, she will send SMS from her mobile (which support GIS) contains her location (Latitude and Longitude) only to the server.

## 6.2 Future work

Improve the proposed system cross Windows Communication Foundation WCF, which can send asynchronous data (messages) from endpoint to another. WCF satisfies rapid communication and other features like [6, 7]:

 *i.* Messages will exchange in one of several patterns.
 *ii.* Security to protect privacy using encryption.
 *iii.* Messages are send on any of several built-in transport protocols and encodings.